\documentclass[sigconf]{acmart}
\AtBeginDocument{%
  }
\usepackage{graphicx}
\usepackage{multirow}
\usepackage{tabularx}
\usepackage{booktabs}
\usepackage{amsmath}
\usepackage{hyperref}
\usepackage{subcaption}
\usepackage{tcolorbox}
\usepackage{bbm}
\usepackage{marvosym}
\setcopyright{acmlicensed}
\copyrightyear{2025}
\acmYear{2026}
\acmDOI{XXXXXXX.XXXXXXX}
\acmConference[]{WWW}{April 13-17,2026}{Dubai, United Arab Emirates}
\acmISBN{978-1-4503-XXXX-X/2018/06}

\begin{document}

\title{From Events to Trending: A Multi-Stage Hotspots Detection Method Based on Generative Query Indexing}


\author{Kaichun Wang\textsuperscript{*},
Yanguang Chen\textsuperscript{*},
Ting Zhang, Mengyao Bao, Keyu Chen, Xu Hu, \and Yongliang Wang,  Jingsheng Yang,  Jinsong Zhang,  Fei Lu}

\thanks{\textsuperscript{*}These authors contributed equally to this work.}

\affiliation{%
  \institution{Bytedance}
  \city{Shanghai}
  \country{China}}
\email{{wangkaichun.16,chenyanguang,zhangting.17,baomengyao,
chenkeyu.526517, huxu.hx}@bytedance.com}
\email{{yongliang.wyl,yangjingsheng,zhangjinsong.jensen}@bytedance.com}

\renewcommand{\shortauthors}{Wang et al.}
\begin{abstract}
LLM-based conversational systems have become a popular gateway for information access, yet most existing chatbots struggle to handle news-related trending queries effectively. To improve user experience, an effective trending query detection method is urgently needed to enable differentiated processing of such target traffic. However, current research on trending detection tailored to the dialogue system scenario remains largely unexplored, and methods designed for traditional search engines often underperform in conversational contexts due to radically distinct query distributions and expression patterns.
To fill this gap, we propose a multi-stage framework for trending detection, which achieves systematic optimization from both offline generation and online identification perspectives. Specifically, our framework first exploits selected hot events to generate index queries, establishing a key bridge between static events and dynamic user queries. It then employs a retrieval matching mechanism for real-time online detection of trending queries, where we introduce a cascaded recall and ranking architecture to balance detection efficiency and accuracy. Furthermore, to better adapt to the practical application scenario, our framework adopts a single-recall module as a cold-start strategy to collect online data for fine-tuning the reranker.
Extensive experiments demonstrate that our framework significantly outperforms baseline methods in both offline evaluations and online A/B tests, and user satisfaction is relatively improved by 27\% in terms of positive-negative feedback ratio.
\end{abstract}

\ccsdesc[300]{Information systems~Trending query detection}
\ccsdesc[300]{Computing methodologies~Natural language generation}

\keywords{Trending Query Detection, Conversational system, Large language model}


\maketitle

\section{Introduction}
With the rapid development of Large Language Models (LLMs) \cite{guo2025deepseek,achiam2023gpt}, AI conversational systems like ChatGPT \cite{ouyang2022training} are gradually becoming a crucial gateway for users to access information. 
Meanwhile, the gradual shift in user information retrieval habits presents new functional demands and practical challenges for conversational applications. Specifically, users sometimes leverage chatbots' online search capabilities to inquire about information on trending events, such as political elections or natural disasters. 
However, current mainstream architectures have obvious shortcomings in handling event-related queries: LLMs themselves lack real-time information, while results returned by general web search suffer from issues such as timeliness delays and information redundancy, making it difficult to ensure response quality.

\begin{figure}[!ht]
    \includegraphics[width=2\columnwidth]{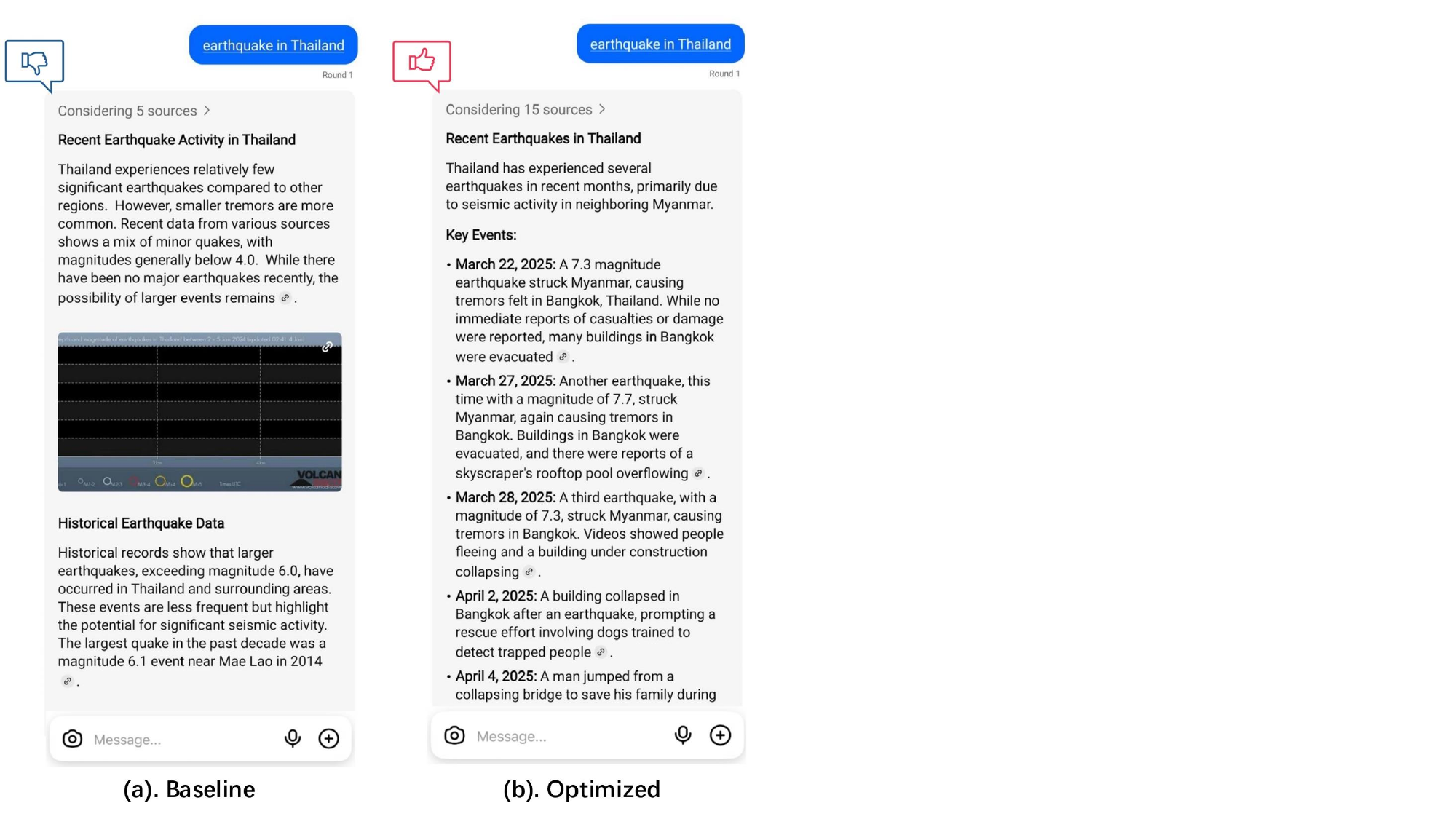}
    \caption{Performance Comparison: Baseline vs. Optimized Pipeline. The baseline supplements its responses with generalized web search results, but the retrieved data lacks timeliness and fails to provide sufficient event information. 
    }
    \label{compare}
\end{figure}

The emergence of Retrieval-Augmented Generation (RAG) \cite{gao2024retrievalaugmentedgenerationlargelanguage} has allowed researchers to directly retrieve supplementary information based on user queries to mitigate the knowledge cutoffs issue of LLMs \cite{vu2023freshllmsrefreshinglargelanguage,hei2024dr}. However, in practical applications, relying solely on RAG to improve the response quality of event-related queries still faces significant challenges. The core lies in the seesaw dilemma between event-related queries and general scenarios: optimizing the entire workflow specifically for trending queries which account for an extremely small proportion will compromise the performance of general scenarios such as knowledge question-answering; conversely, if trending processing adopts the same general workflow without customization, the performance will fail to meet expectations as shown in the Figure ~\ref{compare}. This gives rise to a key research question: \textbf{How to accurately apply event content enhancement to target queries}.

To enhance user experience, an effective event-related trending query detection method tailored to chatbot scenarios is urgently needed, allowing for special handling of these target queries.
In this paper, trending queries are defined as user requests for information about the latest trending events, such as "California wildfires" and "NBA Finals". Such queries are characterized by two key features: they usually lack explicit timeliness keywords, and their content evolves rapidly, making them difficult to classify solely based on text features.

In traditional search engines, researchers usually adopt methods based on log aggregation to identify trending trends through frequency or pattern changes of a large number of similar queries, including time series ~\cite{kulkarni_understanding_2011,shokouhi_detecting_2011} and feature engineering \cite{kanhabua_learning_2015, ghoreishi_predicting_2013, zhang_automatic_2018}.
However, in conversational application scenarios, the sparsity of event-related requests and the diversity of user expressions make it difficult to aggregate queries related to the same event, and hard to capture features such as sudden frequency surges.

To address the aforementioned issues, we propose a multi-stage framework, which systematically delineates the complete technical workflow—from the construction of the event indexing to the accurate identification of trending queries based on generative indexes.

As the data foundation for trending query detection and subsequent processing, our framework first provides a comprehensive solution for \textbf{event database construction}, covering news source selection and event extraction and aggregation. 
Following this, the framework proceeds to the \textbf{trending index generation} stage. For the screened hot events, we generate index queries through prompt engineering, thereby bridging the semantic gap between static event information and dynamically varying user queries. Specifically, to ensure the diversity and reliability of index generation, we introduce a two-stage prompting strategy. First, in the diversity generation stage, we leverage a chain-of-thought (CoT) \cite{wei2023chainofthoughtpromptingelicitsreasoning} strategy to guide the LLM to first reason about the key elements of the event, and then output three types of common sentence patterns to cover the diverse expressions of online users. However, LLMs may generate hallucinations during reasoning, or produce overly generalized sentences in an attempt to comply with diversity instructions. Therefore, we incorporate post filtering for double verification: low-quality indices are eliminated, which not only preserves diversity but also ensures that each index maintains strong relevance to the event content itself, thereby avoiding false retrieval issues caused by low-quality indices.

After generating index queries for trending events, we leverage the semantic similarity between these index queries and user queries to \textbf{detect trending query}. To balance latency and accuracy, we first employ a lightweight embedding model for the retrieval stage; subsequently, we use a generative model to make refined judgments based on more comprehensive information.
Finally, to better adapt the discriminative model to the task, we perform instruction tuning on the pre-trained model using the data collected during the retrieval stage, thereby improving the performance.

In summary, this paper proposes a novel multi-stage framework for trending query detection in AI conversational systems. Starting from hot events, the framework gradually establishes a retrieval-based trending query detection solution through systematic design. Within this framework, our core innovations are mainly reflected in the following aspects:

\begin{itemize}
   \item Tackling multiple challenges in conversational systems, we innovatively recast trending query detection as a retrieval task, which is the first systematic solution to this problem in the domain based on our literature review.

   \item We introduce generative indexing to bridge the semantic gap between static event info and dynamic user queries, balancing index diversity and relevance via CoT and post-filtering.
   
   \item For balancing detection latency and accuracy, we design a collaborative workflow with retrieval and a generative reranker, and use retrieval-stage data to instruction-tune the pre-trained model for better task adaptation.

   \item We validate the effectiveness of our framework through extensive offline experiments and online A/B tests. The results demonstrate that, our framework outperforms baselines significantly in precision-recall and overall system metrics.
\end{itemize}

\begin{figure*}[!ht]
\includegraphics[width=0.97\textwidth]{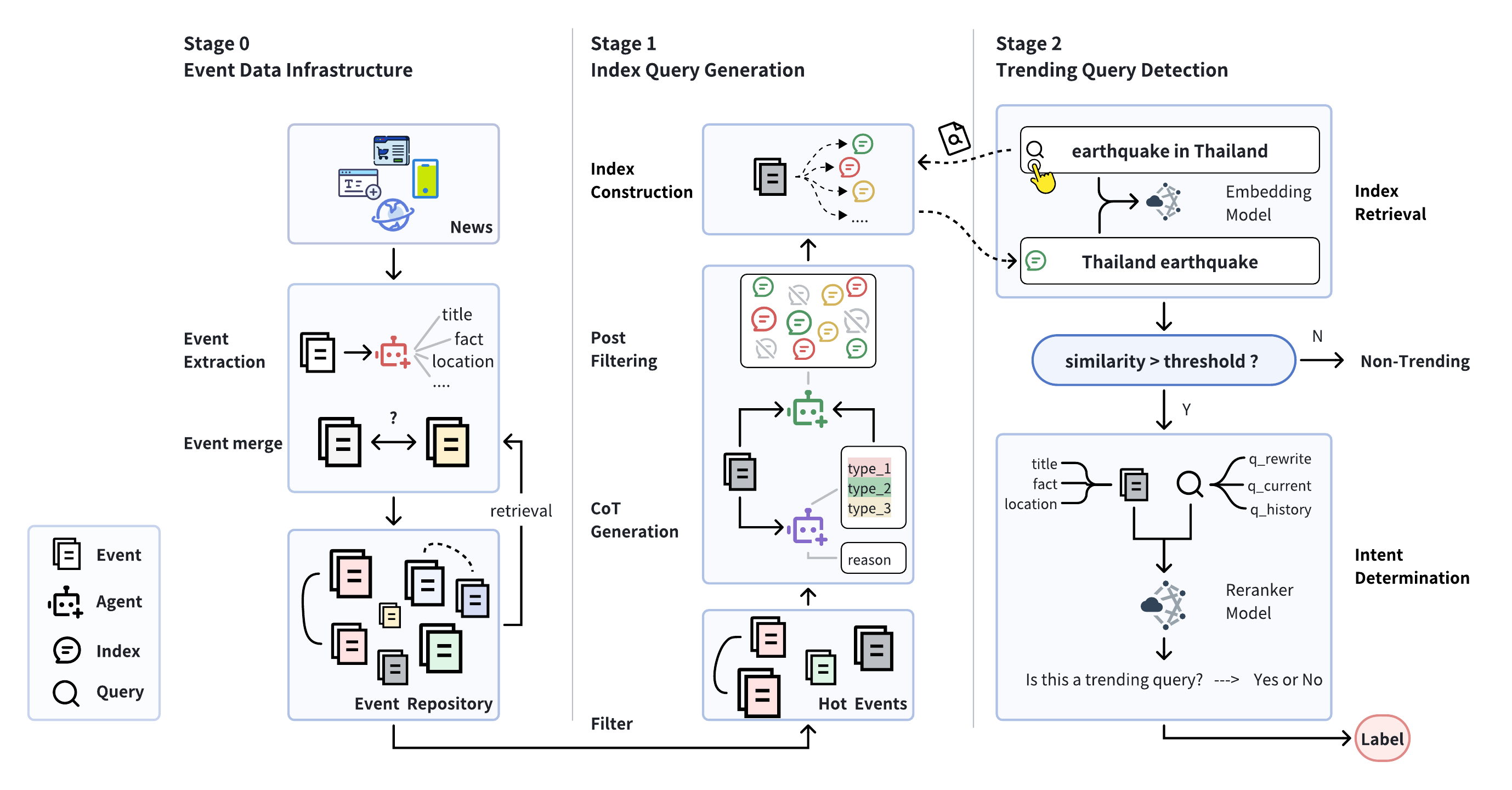}
\caption{An overview of proposed multi-stage framework for trending query detection.}
\label{model}
\end{figure*}

\section{Related Work}
Addressing users' temporal information needs for high-recency content has long been a central challenge in Information Retrieval \cite{piryani_its_2025, zhao_framework_2020}. This challenge becomes particularly salient in conversational search, which requires systems to comprehend not only the literal meaning of a query but also the user's implicit intent toward recent events. Existing research offers valuable insights but also reveals significant limitations, which we categorize into three methodological strands.


\subsection{Query-Centric Temporal Modeling}

Early research primarily focused on identifying temporal queries through analysis of search logs. These methods can be broadly divided into two categories: those based on temporal patterns in query popularity and those based on feature engineering.

\subsubsection{Time-Series Analysis of Query Popularity.\\}
A line of work detects temporal sensitivity by analyzing time-series signals in query volume. These approaches identify characteristic patterns—such as spikiness, periodicity, or abrupt shifts in popularity—to determine whether a query is event-sensitive or recurrent \cite{kulkarni_understanding_2011, shokouhi_detecting_2011, mansouri_detecting_2017}. The underlying assumption is that queries related to breaking events exhibit distinctive temporal signatures distinguishable from those of perennial queries.

\subsubsection{Feature-Based Classification.\\}
Another strand employs supervised learning with carefully engineered features to predict a query's event-relatedness. These classifiers typically integrate multiple feature types, including temporal patterns, click-through rates, and content-based characteristics \cite{kanhabua_learning_2015, ghoreishi_predicting_2013, zhang_automatic_2018}. By combining these signals, the models learn to distinguish queries seeking recent information from those with more stable information needs.

A fundamental limitation of both approaches is their heavy dependence on abundant historical query logs, which renders them ineffective for emerging events with little or no prior data. More critically, they operate primarily at a statistical level and lack deep semantic understanding, often failing to disambiguate implicit intents—for example, determining whether a query containing a person's name seeks biographical information or breaking news.

\subsection{Event-Centric Content Organization}

To address the limitations of query-only methods, subsequent work organized content around structured event representation. A common strategy is to cluster news articles or social media posts into event-centric units using metadata such as titles, entities, and publication times \cite{gupta_eventminer_2016}. A notable example is Event-centric Query Suggestion (EQS) \cite{sachin_event-centric_2022}, which constructs event clusters from news metadata and suggests queries based on keyword matching.

While EQS represents progress toward event-aware retrieval, it faces two critical shortcomings. First, its event construction relies heavily on consistent and high-quality metadata, which is often unavailable in practice. Second, its query matching is based on lexical overlap, making it vulnerable to vocabulary mismatch and unable to handle paraphrases or implicit intents \cite{mansouri_learning_2017}. These issues highlight the need for richer semantic representations of events and more robust intent matching.

\subsection{Limitations in the LLM Era}

The rise of LLMs and conversational search has further exposed the inadequacies of earlier methods. Modern interactive systems require human-like comprehension and context awareness, going far beyond keyword matching \cite{mo_survey_2025, yuan_query_2025}. In conversational settings, misinterpreting a general query as a news request can severely degrade user experience—a risk not fully addressed by prior work.

Moreover, LLMs are known to suffer from knowledge cutoffs, prompting a growing emphasis on connecting them with real-time information sources \cite{noauthor_llms_nodate, bytezcom_are_2025, conlen_we_2024}. While this direction is widely recognized, few frameworks systematically integrate real-time content crawling, event abstraction, and LLM-powered intent detection into a unified pipeline. Our work bridges this gap by not only leveraging LLMs for semantic understanding but also deploying them to link dynamic user intents with timely, event-grounded content—enabling accurate, intent-aware news retrieval even for long-tail events.

\section{Method}

\subsection{Problem Formulation}
Our task is to determine whether a user's search query is a trending query related to an ongoing hot event. This decision is made by integrating multi-dimensional information  within a conversational context and based on the hot event information we filtered. 
The input query is a composite representation $Q$, which integrates three distinct components: the rewritten query generated by the model $q_{r}$ (which summarizes information from the user's recent three dialogue turns), the original user query from the current turn $q_o$, and the historical queries from the recent two dialogue turns $q_h$. Formally, we represent the user query as $Q(q_r, q_o, q_h)$. 
Instead of traditional binary classification, we reframe it as a cascaded process of retrieval and discrimination. In the retrieval stage, given the user’s rewritten query $q_r$ and the index query set $\mathcal{I}$, the model identifies the nearest neighbors of the user’s query via semantic similarity matching. Queries with similarity meeting the threshold condition proceed to the discrimination stage, where the reranker model determines the final label based on the query $Q$ and event information $E$.

This cascaded paradigm effectively combines the recall advantage of efficient semantic retrieval with the precision of powerful generative discriminators, enabling our system to robustly identify event-related trending queries from user conversations.

\subsection{System Architecture}
The architecture of our framework, illustrated in Figure \ref{model}, details the comprehensive pipeline. The pipeline consists of the following key phases:

\begin{enumerate}
   \item \textbf{Events Database Construction}: The first stage of our framework offers a comprehensive solution for building events database, covering news source selection, event extraction and merge. This database acts as a critical foundation for subsequent trending query detection and response.
   
   \item \textbf{Index Query Generation}: For each selected hot event, we use a LLM to generate a diverse set of search queries as indexes, while leveraging constraints and filtering strategies to ensure these queries are of high quality and relevance.

   \item \textbf{Trending Query Detection}: Based on the generated index queries, we detect trending queries by cascading retrieval with a fine-tuned reranker model. This approach maximizes the system’s detection accuracy within latency constraints.
\end{enumerate}

\subsection{Events Database Construction}

\begin{figure*}[htbp]
\includegraphics[width=0.9\textwidth]{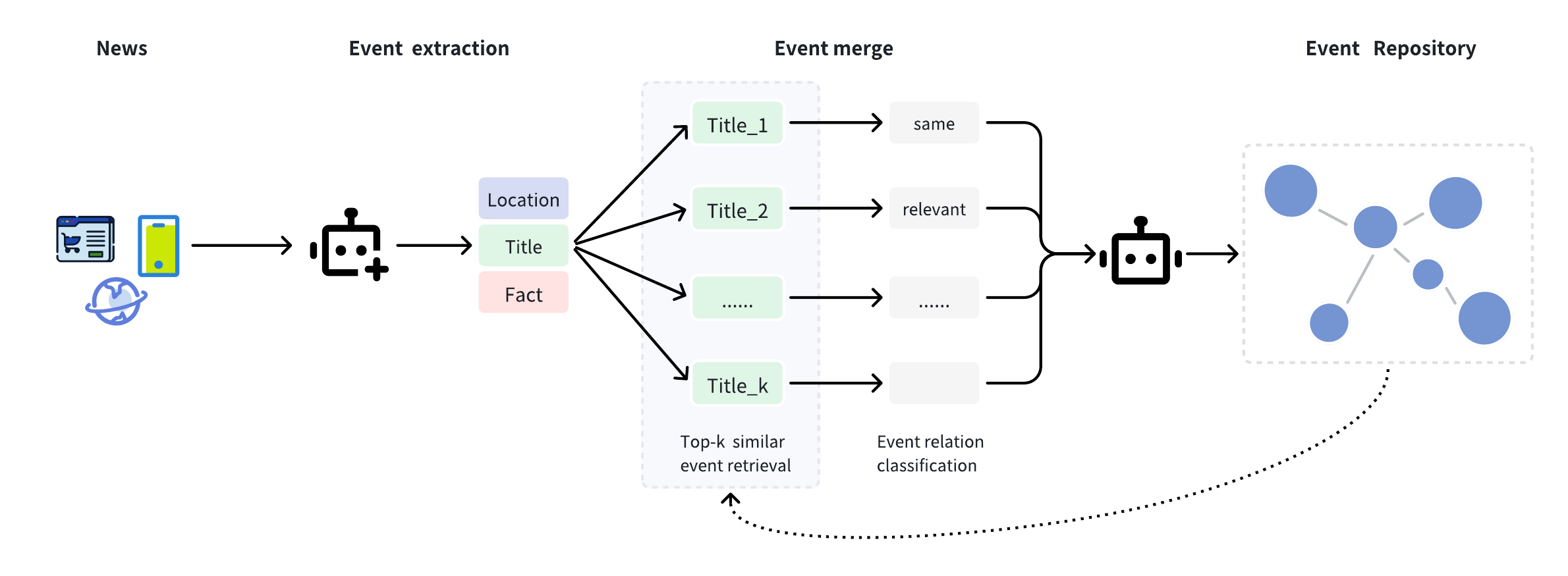}
\caption{Event repository construction pipeline.}
\label{event}
\end{figure*}

In this section, we introduce the event repository construction process that serves as the foundation for trending query generation. As shown in Figure \ref{event}, our approach addresses two critical aspects: (1) strategic selection of authoritative news sources to ensure data quality and trending potential, and (2) systematic event extraction and construction to facilitate efficient query generation. This dual-focus methodology ensures both the effectiveness of the resulting trending queries and the scalability of our overall system.

\textbf{News Source Filtering.} The first component of our methodology focuses on establishing a high-quality news source foundation. We implement a systematic curation process by selecting high-ranking news and magazine websites from both mobile applications and web platforms across multiple countries. This selection strategy ensures comprehensive coverage of international and local newsworthy events spanning diverse domains including technology, sports, finance, and entertainment. To maintain data quality and authenticity, we filter out non-authoritative sources based on website popularity and reputation, while restricting our selection to websites that comply with legal regulations and permit automated data collection. This systematic curation process yields a robust dataset of internationally and locally relevant news events that serve as reliable candidates for trending query identification.

\textbf{Event Extraction and Consolidation.} While the curated news sources provide high-quality content, directly generating trending queries from raw news articles presents significant efficiency challenges. Individual news articles exhibit inherent limitations that impede effective query generation: content fragmentation, redundant coverage of the same trending topics, inconsistent article lengths, and varying information density. More critically, generating queries directly from thousands of individual articles would be computationally prohibitive and likely produce numerous duplicate or low-quality queries focusing on the same underlying events.

To address these challenges, we implement a two-stage processing pipeline that transforms raw articles into consolidated event representations. 
During the extraction stage, we systematically retrieve structured information from individual news articles. 
Given a news report, we will extract a comprehensive event structure, including factual information such as the title, event location, and key outcomes.
Using LLMs and well-designed prompts, we extract three key components—title, time, and factual content—from each article. To avoid LLM hallucination-related issues, we only ask the LLM to extract factual content from the original text during extraction, with no further processing or divergent expressions allowed.


Following extraction, the consolidation stage identifies and merges events, which describe the same news event, to eliminate redundancy and enhance information density. We first retrieve top-k candidate events based on BGE-M3 \cite{chen2024bge} similarity scores of abstracts. For event merging, we fine-tune the InternLM2-1.8B \cite{Cai2024InternLM2TR} model to create a specialized event relationship classification model. This model performs pairwise relationship assessment between incoming events and candidate events in the event repository. Related events identified as describing the same underlying occurrence are systematically merged using other generative model, resulting in comprehensive, non-redundant event representations that serve as high-quality foundations for trending query generation.

\subsection{Event Index Generation}
This section describes the process of generating index queries, which consists of two main components: hot event selection and query generation. A systematic scoring mechanism is designed to identify influential and timely events from the event repository. For each selected event, a diversified set of search queries is generated using a LLM, with constraints and filtering strategies applied to ensure high quality, relevance, and coverage for retrieval.

\subsubsection{Hot Event Selection.}
We first filter the event information within the afore described event repository to select the hot events that will serve as the data source for the generative trending queries. We define "hot events" as those exhibiting great influence, high exposure and strong timeliness, informed by operational experience from contemporary news recommendation systems.

To quantify the effect of influence, we train a scoring model to produce a model score $S_m$ for each event by curating a set of typical instances and edge cases spanning different levels of the attribute and collecting crowd-sourced annotations. We introduce a hyperparameter domain weight $W_{d}$ to effect a fine-grained, domain-specific weight calibration of the scores, considering that the social attention varies across verticals (e.g., political news and sport news generally elicit higher discussion than automotive or consumer electronics news). As a proxy for exposure, we use news count $C_n$ which represents the number of news articles aggregated under a event. Furthermore, we apply time decay with respect to the event timestamp to capture temporal dynamics.

The final score for a hot event is computed as:
\begin{equation}
Score 
= 
S_m W_{d}(1+ \lambda_c C_n)(1+\alpha_{t}*0.5^{\Delta t/21600})  ,
\end{equation}
where $\alpha_t$ controls the magnitude of the temporal term, $\Delta t$ represents the time difference between the current time and the event occurrence time, and the exponential factor implements a half-life of six hours. Events with a score > 0.5 are designated as hot events and are retained for downstream index construction.



\subsubsection{Index Query Generation.}

We leverage GPT-4o-mini to generate multiple hotspot queries which are highly pertinent to a given event. The inputs comprise basic event metadata, including the event title,  fact and locations. The generated hotspot queries collectively cover salient entities and key facts of the event from various perspectives.

We evaluate two primary dimensions, diversity and relevance. Diversity characterizes the variation among the queries generated from a single event. Since user queries vary in both expression and information-seeking perspective, the generated queries must be diverse and closely resemble natural user phrasing to ensure broad coverage during the retrieval stage. Relevance measures the alignment between each trending query and the current event information. It reflects whether a query indeed targets the current hot event, thereby supporting precision in the retrieval stage.

For the purpose of achieving these objectives, we meticulously design our prompts. Drawing on the Chain-of-Thought (CoT) paradigm, we first extract the salient entities from each event, which serve as the foundation for generating queries. We then design prompts to elicit three distinct sentence structures and different permutations of the entities to increase the diversity of the generated outputs. To improve relevance, we guide generation focusing on salient entities and actions of the event. We further adopt a few-shot prompting strategy to steer the model toward producing highly pertinent and information-rich queries. Concurrently, we periodically analyze failure cases observed in the online retrieval stage and iteratively refine the prompts.
We apply an LLM-based post filtering procedure to further improve hotspot query quality. We input the event information together with its generated hotspot queries and filter out queries with missing or incorrect information by leveraging the model’s text understanding. Finally, we establish a strict timeout window by coupling the event lastest update timestamp with the hotspot queries. Only queries associated with events updated within this window remain active. The timeout is set to 7 days by default.

\subsection{Trending Query Detection}

After generating index queries for trending events, we leverage their semantic proximity to actual user queries to enable online identification of trending queries. The process of online trending query identification consists of two main steps: First, we employ semantic retrieval to recall the top-K event indices from the generated trending event index that are semantically similar to the current user query. Second, we retrieve the original event information corresponding to these K indices and provide both the event information and the user's recent dialogue history to the LLM, which determines whether the user is inquiring about a specific trending event.
In practical applications, we set K to 1.

\subsubsection{Event Index Retrieval.}

The primary objective of trending index retrieval is to identify semantically similar content from the pool of generated index queries. We observe that user queries in conversational scenarios tend to be more flexible and diverse in their expression. Therefore, instead of using only the most recent user query for retrieval, we rewrite the last three rounds of user queries to construct a more comprehensive intent query for retrieval. To ensure efficient online retrieval, we utilize embedding-based methods. Specifically, we employ the pre-trained BGE-m3 model to compute embedding vectors for both the rewritten user query and the index queries, and then perform top-K retrieval based on vector inner product similarity. Given that queries with trending intent constitute a very small fraction of all online queries (approximately 0.5\%), a significant portion of the recalled index queries are of low relevance. Thus, it is essential to filter out queries that are clearly unrelated to trending topics. Through random sampling and evaluation of online traffic, we select the inner product threshold that maximizes the F1 score. Only queries with an inner product above this threshold are considered valid recalls, effectively reducing the volume of non-trending traffic entering the precise intent determination stage.

\subsubsection{Trending Intent Determination.}
For trending queries retained after threshold filtering, a more precise intent determination is required. Since the rewritten queries used for retrieval may still introduce ambiguity, we directly use the user's most recent three queries as input at this stage. On the content side, rather than relying solely on the generated index queries, we retrieve the original event information corresponding to the recalled indices. Both the original event information and the recalled indices are provided as input to the LLM, which assesses whether the user's query intent is to inquire about information related to a trending event.

We constructed a query-document evaluation dataset by manually annotating real user queries and trending events. On this dataset, we compared the performance of various open-source models in relevance matching. Based on these comparisons, we selected beg-gemma as the base model for relevance matching. We further conducted comparative studies on the event content used for matching, including using only the generated index, only the event title, and the original event information. Our results indicate that the original event information yields the best performance, as it maximizes the information content of trending topics.

Based on the conclusions of the comparative experiment, we selected query-event data pairs as input to fine-tune the bge-gemma reranker model.
Notably, a large portion of the queries sampled in the retrieval stage are false positives, with no corresponding hot events as positive samples, so we cannot adopt the common pairwise method based on contrastive learning. 
To adapt to the online sampled data, we modified the original fine-tuning interface of BGE and trained the model using the pointwise method. We optimize our model by minimizing the Cross-Entropy loss $\mathcal{L}$:
\begin{equation}
\mathcal{L} = - \left[ y \cdot \log(\hat{y}) + (1 - y) \cdot \log(1 - \hat{y}) \right], 
\end{equation}
where $y$ represents the true label of the sample, and $\hat{y}$ represents the predicted probability of the model output by $Sigmoid$.


\section{Experiments}
\subsection{Setup}
Our framework is implemented and evaluated in a large-scale production-level conversational AI system. To guarantee that the models are trained and tested using representative data, all datasets are sampled from anonymized real-world user interactions with the live system.
In the retrieval stage, we adopt BGE-M3 \cite{chen2024bge} as the embedding model for its state-of-the-art performance and computational efficiency——two key requirements for deployment in a production environment. For the reranker model, we selected three high-performance models for comparison on the same test dataset, and ultimately chose the bge-reranker-v2-gemma.

During the online deployment phase, to ensure the service remains stable even during peak traffic periods, we have configured 8 A10 GPUs to support the Embedding service in the retrieval stage, and additionally deployed 4 A30 GPUs to back the Reranker model service in the fine-grained discrimination stage.

\subsubsection{Datasets.}
Our dataset comprises two categories: test sets for offline evaluation of the retrieval and discrimination stages, and a training set for fine-tuning the reranker model. The details are as follows:

\begin{itemize}
   \item \textbf{Retrieval-stage test dataset}: With a 3-day cycle for hotspot events, 70 hotspot events from two cycles were selected as candidates. 12,000 online queries were randomly sampled and manually labeled to evaluate the detection performance.
   \item \textbf{Reranking-stage test dataset}: : 2,000 online query-event pairs output from the retrieval stage were manually labeled and used as the evaluation set for the reranking module. Notably, since all samples have been filtered by the pre-retrieval module, the negative samples are all hard cases, posing a significant challenge to the reranker model.
   \item \textbf{Fine-tuning dataset for the reranker model}: To meet task requirements, 13,000 online data samples from 9 days were collected. Labels were finalized through auxiliary labeling by LLMs and manual review, which were used for fine-tuning the pre-trained reranker model.
\end{itemize}

\subsubsection{Metrics.}
Our evaluation is divided into two forms: online and offline, and the corresponding evaluation metrics also differ.

\textbf{Online evaluation.}
To assess the real-world performance of the strategies we proposed, we conducted a 7-day online A/B test on our conversational AI platform. In this experiment, each trending query detection strategy was deployed as a unique variant and allocated 40\% of the total user traffic for evaluation.
The core metrics included two key indicators: the platform's last\_3\_day user activity (referred to as "user activity" for short) and the positive-to-negative feedback ratio under hotspot traffic (referred to as "user satisfaction" for short). In addition, we selected traffic coverage rate and the actual accuracy calculated via online sampled labeling to measure the precision and recall performance of the methods.

\textbf{Offline evaluation.}
To explore the optimal solutions for the retrieval and discrimination modules in depth, we conducted experimental evaluations on the two modules separately during the offline phase. During the evaluation, the main focus was on the recall and precision of each individual module on the corresponding evaluation dataset.
It should be noted that recall and precision fluctuate with changes in threshold settings. Therefore, for each experimental configuration, we selected the threshold parameter that yielded the highest F1-score to measure the optimal comprehensive performance of the module under that configuration.

\subsection{Main Results}
In this section, we present the experimental results for different trending query detection strategies and methods for each stage.

\begin{table}[!ht]
  \centering
  \caption{Results of different strategies for online A/B testing, where "2 stage" represents the cascaded solution of retrieval and discrimination.}
  \scalebox{0.9}{
  \begin{tabular}{lcccc}
    \toprule
    \textbf{Methods} & \textbf{Activity gain} & \textbf{Satisfaction gain} & \textbf{coverage}  & \textbf{precision} \\
    \midrule
    Baseline           & +0\% & +0\% &  0\% & \textbackslash \\
    retrieval    & +0.38\% & +13.47\% &  0.37\%  & 0.72 \\
    2 stage            & +0.77\% & +21.73\% &  0.50\% & 0.87 \\
    2 stage\textsubscript{ft}  & +0.91\% & +27.65\% &  0.51\% & 0.93 \\   
    \bottomrule
  \end{tabular}
  }
  \label{tab:strategy_metrics}
\end{table}

\subsubsection{End to End A/B Test.}

The accuracy of trending query detection is not an isolated technical indicator, it directly determines the quality of user interaction with conversational systems, and further exerts a significant indirect impact on online core metrics (e.g., user activity) and user satisfaction. The transmission mechanism of this impact can be specifically explained through the contradictions in two typical scenarios:

{Response deficiencies for real trending queries.}
For queries related to real hotspots (see Figure 1), the traditional baseline pipeline has inherent shortcomings. Since it does not integrate real-time information and structured content of hot events, and only generates responses based on general dialogue logic, it often fails to meet users’ core needs for timeliness and completeness of hotspots. 

{Misjudgment of non-trending queries.}
If a non-trending query is misjudged as trending, it activates the news event processing pipeline, causing new interaction problems. The event pipeline prioritizes presenting event information (its responses focus on events), but non-trending queries (e.g., "how to cook rice") often relate to daily consulting. Misclassification here may force the system to return rice-related hotspots (e.g., a brand’s quality issue), missing user needs. This frustrates users, increases negative feedback, and reduces satisfaction.

Therefore, in the online A/B test evaluation of each of our strategies, we focused on two key aspects: detection accuracy and the positive impact on core online metrics.
The experimental results for our proposed framework are presented in Table \ref{tab:strategy_metrics}. The results demonstrate a consistent and significant improvement across all metrics as we progress through the stages of our pipeline.
Key findings are highlighted below.

\textbf{Gains and limitations of single retrieval module.}
Compared with the non-event-optimized baseline, simply introducing a retrieval module (index retrieval) effectively identifies hot - spot queries and brings significant gains. With trending query traffic accounting for just 0.3\% of total traffic, it boosts overall user activity by 0.38\%, and user satisfaction with responses rises by over 13\%. However, this strategy faces a precision-recall trade-off. To ensure online judgment precision is no less than 0.7, a relatively high matching threshold must be set, causing some real trending queries to go unrecalled and leading to insufficient coverage.

\textbf{Optimization of retrieval by reranking Module.}
Introducing a reranking model based on the retrieval module effectively alleviates the trade-off between precision and recall. Leveraging the reranking model’s stronger semantic discrimination capability, the threshold in the retrieval stage can be appropriately lowered to improve recall, while more accurate judgment is delegated to the reranking stage. This improvement not only enhances both traffic coverage and online precision but also indirectly boosts user activity and satisfaction—with user activity increasing by 0.39\%. Notably, the introduction of the reranker has led to a significant 0.15 increase in precision, and traffic coverage has risen by approximately 0.14\% compared to the previous version of the strategy.

\textbf{Additional benefits of fine-tuning reranking model.}
After instruction tuning of the reranking model, the online precision improves by approximately 0.06, without sacrificing recall performance (traffic coverage remains nearly unchanged). The enhanced precision indirectly boosts user experience: compared with the non-tuned strategy, the overall user activity increases by an additional 0.39\%, and user satisfaction rises by 5.92\%, achieving simultaneous optimization of both precision and user experience.

\begin{table}[!ht]
  \centering
  \caption{Performance of different retrieval  methods. }
  \begin{tabular}{lcc}
    \toprule
    \textbf{Method} & \textbf{Recall} & \textbf{Precision} \\
    \midrule
    event\_title           & 0.64            & 0.70               \\
    event\_fact           & 0.67            & 0.71               \\
    gen\_index           & 0.78            & 0.81               \\
    gen\_index\_diversity & 0.83            & 0.88               \\
    gen\_index\_filter    &  \textbf{0.83}    & \textbf{0.90}    \\
    \bottomrule
  \end{tabular}
  \label{recall}
\end{table}

\subsubsection{Retrieval Methods Evaluation.}
Next, we explore the performance of different methods in the retrieval module. To verify the effectiveness of generative indexes, we compared two approaches: one using original event information as the retrieval object, and the other using index queries generated by LLMs based on event content.

From the experimental results presented in Table ~\ref{recall}, we can infer the following conclusions: 

\textbf{Generative indexes outperform event-based retrieval in effectiveness.}
As the content of hot events is fixed, while user queries for such events in conversational scenarios vary significantly, relying solely on original event features (titles or facts) fails to cover the diverse expression of user needs. By contrast, generative index address this limitation: Table ~\ref{recall} shows gen\_index outperforming event\_title and event\_fact notably. This confirms that generative index effectively expand the coverage of user query expressions, leading to a substantial improvement in retrieval Recall.

\textbf{CoT-guided diversity enhancement improves both Recall and Precision of generative index.}
To enhance the diversity of generative indexes, we adopted a CoT strategy: guiding the model to first analyze the key content of hotspot events, then generate queries in three distinct sentence structures. This optimization boosts the diversity of index expressions, enabling the retrieval system to match more varied user queries—reflected in a higher Recall of gen\_index\_diversity (5 percentage points higher than the basic gen\_index). Meanwhile, the improved index diversity allows for an appropriate increase in matching thresholds, which reduces false positive matches and elevates Precision, realizing simultaneous improvements in both core metrics.

\textbf{LLM-based post-filtering ensures index relevance while maintaining diversity.}
During CoT-guided diverse index generation, LLMs may produce hallucinations or deviate from hot event content in pursuit of excessive diversity. To mitigate this issue, we introduced an LLM-based post-filtering mechanism for dual verification: filtering out ambiguous, low-quality index queries that lack strong relevance to hot events. As shown in Table ~\ref{recall}, gen\_index\_filter retains the high Recall while further improving Precision to 0.90. This demonstrates that post-filtering effectively balances diversity and relevance, ensuring each index maintains a strong association with the target hot event.

\begin{table}[ht]
  \centering
  \caption{Performance of Different Methods}
  \begin{tabular}{llcc}
    \toprule
    \textbf{Category} & \textbf{Method} & \textbf{Recall} & \textbf{Precision} \\
    \midrule
    \multirow{3}{*}{\textbf{Model}} 
    & bge-m3-reranker~\cite{chen2024bge}       & 0.79 & 0.71 \\
    & bge-gemma~\cite{chen2024bge}             & 0.78 & 0.82 \\
    & qwen3-reranker-4b~\cite{qwen3embedding}  & 0.70 & 0.64 \\
    \midrule
    \multirow{3}{*}{\textbf{Feature}} 
    & query - index                             & 0.74 & 0.73 \\
    & query - title                             & 0.69 & 0.72 \\
    & query - event                             & 0.79 & 0.82 \\
    \midrule
    \textbf{Optimized} 
    & bge-gemma-sft                             & \textbf{0.90} & \textbf{0.92} \\
    \bottomrule
  \end{tabular}
  \label{tab:method_performance_2}
\end{table}

\subsubsection{Determination Methods Evaluation.}
For the discrimination module, we conducted experiments on different base models and input feature combinations, with the results shown in Table ~\ref{tab:method_performance_2}.

We can observed that among the evaluated baseline models, bge-gemma achieves the most balanced and competitive performance. This confirms that bge-gemma is the optimal choice among the tested baseline models for subsequent discriminative tasks.

To optimize the input features for bge-gemma's discriminative task, we compared three feature combinations. The results show that the "query-event" combination (using user queries and full hot event information) outperforms the other two options. The superior performance of 'query-event' is attributed to the comprehensive information provided by full hot event details, which enables the model to capture more contextual and semantic connections between user queries and target events during reasoning.

Finally, building on the above conclusions, we further fine-tuned the bge-gemma model using "query-event" pairs as training data. The resulting optimized model achieves significant performance improvements: its Recall jumps to 0.90 and Precision rises to 0.92, far exceeding both the original bge-gemma and the "query - event" feature-based discriminative results. This fine-tuned model thus serves as the final reranker deployed in the online system.





\subsection{Ablation Study}
In this section, we conduct ablation studies to validate the effective-ness of our key strategies, focusing on index query generation and reranker SFT.

\begin{table}[!ht]
  \centering
  \caption{Ablation on indexed query pattern. }
  \begin{tabular}{lc}
    \toprule
    \textbf{Method} & \textbf{F1-score}  \\
    \midrule
    gen\_index\_diversity & 0.86                    \\
    w/o reason     & 0.80      \\
    w/o factual    &  0.81     \\
    w/o search     & 0.84      \\
    w/o question     & 0.82      \\
    \bottomrule
  \end{tabular}
  \label{index}
\end{table}

\textbf{Ablation on indexed query pattern.}
As mentioned earlier, when generating index queries, we enforce the creation of three sentence patterns: factual queries, search queries, and question queries, with specific definitions provided in the appendix.

To explore the role of each factor, we attempted to ablate both the intermediate reasoning steps of CoT and each sentence pattern, the results are shown in Table \ref{index}.
We found that in diverse generation, each optimization factor influences to varying degrees. Notably, among all sentence patterns, factual description queries are most crucial, partly because they cover multi-dimensional event information and users tend to search information more through factual statements.

\begin{table}[!ht]
  \centering
  \caption{Ablation on SFT data, where "dayn" denotes sampling data from n days for SFT. }
  \begin{tabular}{lcc}
    \toprule
    \textbf{Method} & \textbf{Recall} & \textbf{Precision} \\
    \midrule
    day9           & 0.90            & 0.92               \\
    day4           & 0.87            & 0.90               \\
    day3           & 0.85            & 0.89               \\
    day1           & 0.84            & 0.87               \\
    \bottomrule
  \end{tabular}
  \label{SFT}
\end{table}

\textbf{Ablation on SFT data.}
Due to the periodicity of hot events, we collected nine days of online data to fine-tune the base model that ultimately serves online scenarios. To verify how data from different time periods affects model fine-tuning, we conducted an ablation study on the data.

As presented in Table \ref{SFT}, which reports the results of the ablation study on SFT data, we can observe a clear trend regarding the impact of data sampled over different durations on model fine-tuning. Essentially, as the number of sampling days increases, the model can learn more patterns of associations between users' queries and event information. This helps enhance the model's generalization ability, making it better suited to the task scenario.

\section{Conclution}
In this work, we propose a multi-stage framework for conversational trending query detection, which integrates events database construction, event index generation, and trending intent detection. 
The proposed framework  transforms static classification and recognition into a dynamic retrieval task. It adapts to the diverse expressions of users in conversational systems through generative indexes, filling the research gap in this task within the field.
Our framework achieved an F1-score of up to 0.91 in offline evaluation; more importantly, it increased user satisfaction by over 27\% in online scenarios.

\clearpage
\bibliographystyle{ACM-Reference-Format}
\bibliography{event,model}


\newpage

\appendix

\section{Prompt Template}

\begin{figure*}[!ht]
\includegraphics[width=\textwidth]{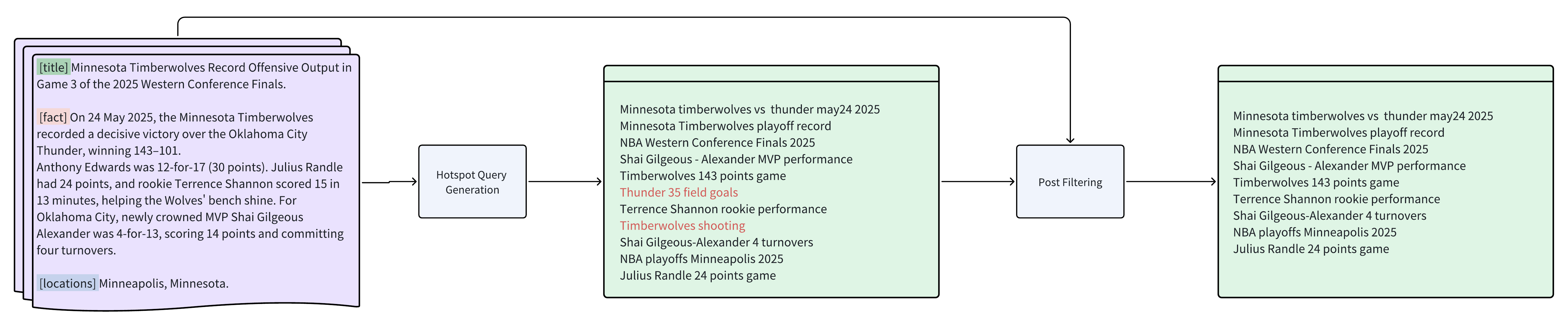}
\caption{Event repository construction pipeline.}
\label{example}
\end{figure*}

\subsection{Index Query Generation}
In the index query generation stage, we include two phases: diverse generation and post filtering, an example is shown in the Figure \ref{example}.
First, the prompt for the diverse generation phase is as follows:

\begin{center}
\begin{tcolorbox}[colback=gray!10,
                  colframe=black,
                  width=\linewidth,
                  arc=1mm, auto outer arc,
                  boxrule=0.5pt,
                 ]

\#\# \textbf{Role} 

You’re a news hotspot editor tasked with organizing news hotspot event descriptions. You’ll receive news event info and need to rewrite it into short sentences matching ordinary users’ search habits.

The news event description is: \textbf{\{\{event\_info\}\}}

\textbf{\#\#Workflow}

1. Read and understand the event description, identify the core information and important entities in the event.

2. \textbf{Factual query: }Centered on the key information in the event description and the judgment results, rewrite 30 factual phrases with different sentence structures that align with users' search habits. Ensure diverse sentence structures, no repeated information, and that the provided factual phrases can accurately search for this event.

3. \textbf{Search query: }Based on the results of the factual phrases, rewrite 10 search-style phrases. All of them must be interrogative sentences, strictly align with user habits, and have diverse sentence structures.

4. \textbf{Question query: }Based on the results of the factual phrases, rewrite 6 question-type phrases that align with users' search habits.\\

\textbf{\#\# Output Format}

\textbf{\{Reason\}} \\
\textbf{\{Factual queries\}} \\
\textbf{\{Search queries\}} \\
\textbf{\{Question queries\}} 
\end{tcolorbox}
\end{center}

Subsequently, in the post filtering phase, we take the event information and the queries generated in the previous phase as input, and let the LLM judge their relevance. The prompt is as follows:

\begin{center}
\begin{tcolorbox}[colback=gray!10,
                  colframe=black,
                  width=\linewidth,
                  arc=1mm, auto outer arc,
                  boxrule=0.5pt,
                 ]

\textbf{\#\# Role} 

You will receive news event description info and its corresponding lists of English factual, search-style, and interrogative phrases. Filter out low-quality ones, select usable phrases, and return them.

The news event description is: \textbf{\{\{event\_info\}\}} \\
\textbf{\{\{Factual queries\}\}} \\
\textbf{\{\{Search queries\}\}} \\
\textbf{\{\{Question queries\}\}}

\#\# \textbf{Workflow}

Assess the quality of the factual phrases according to the \textbf{\{\{Quality Standard\}\}}.\\
...

\#\# \textbf{Output Format}

\textbf{\{\{Filter queries\}\}} 
\end{tcolorbox}
\end{center}

In the filtering process, we expect the LLM to adhere to the following quality criteria.

\begin{itemize}
   \item Use the reference event as background information to ensure the content of the phrases is authentic.

   \item Phrases must be relevant to the event; phrases with insufficient relevance are unqualified.
   
   \item Phrases must not lack specific locations, specific entities, or key information related to the event.

   \item Phrases must not contain words irrelevant to the event.
\end{itemize}

\subsection{PE for Reranker Model}
To better adapt to our task scenario, we used the following prompt as input during the fine-tuning phase.

\begin{center}
\begin{tcolorbox}[colback=gray!10,
                  colframe=black,
                  width=\linewidth,
                  arc=1mm, auto outer arc,
                  boxrule=0.5pt,
                 ]
Given query(A)-passage(B) pair, execute sequential checks  by providing a prediction of either 'Yes' or 'No':\\

1. Intent Validation\\

IF A lacks discernible intent purpose (e.g., is empty or contains only whitespace) → 'No'

IF A uses vague patterns (e.g., "anything about...", "something related to...") → 'No'

IF A has a clear and specific intent and does not use vague patterns → Continue to next check\\

2. Entity Grounding Test\\

IF A does not contain at least one named entity from [Person] | [Organization] | [Domain - Specific Concept] → 'No'

IF A only contains generic nouns (e.g., "something good", "a thing") → 'No'

IF A contains at least one named entity from the specified categories and no only - generic nouns → Continue to next check\\

3. Content Verification\\

IF A is a refined query with specific details and constraints in addition to the key named entity AND B contains all the key information from A → 'Yes'

IF A simply mentions a clear key named entity without many additional details AND B provides sufficient information centered around that entity → 'Yes'

In all other non - passing cases of Content Verification → 'No'
\end{tcolorbox}
\end{center}

\end{document}